\documentclass{article}
\usepackage{ijcai22}

\usepackage{times}

\usepackage{soul}
\usepackage{url}
\usepackage[hidelinks]{hyperref}
\usepackage[utf8]{inputenc}
\usepackage[small]{caption}
\usepackage{graphicx}
\usepackage{amsmath}
\usepackage{amsfonts}
\usepackage{enumitem}
\usepackage{booktabs}
\usepackage{xcolor}
\usepackage{subfig}

\usepackage[acronym]{glossaries}
\makeglossaries

\newacronym{vae}{VAE}{Variational Auto-Encoder}
\newacronym{ae}{AE}{Auto-Encoder}
\newacronym{am}{AM}{Attention Mechanism}
\newacronym{tf}{TF}{Transformer}
\newacronym{gan}{GAN}{Generative Adversarial Network}
\newacronym{cl}{CL}{Correlation Learning}
\newacronym{avcl}{AVCL}{Audio-Visual Correlation Learning}
\newacronym{ml}{ML}{Machine Learning}

\newacronym{al}{AL}{Adversarial Learning}
\newacronym{mlp}{MLP}{Multilayer Perceptrons}
\newacronym{cnn}{CNN}{Convolutional Neural Network}
\newacronym{dnn}{DNN}{Deep Neural Networks}
\newacronym{mmml}{MMML}{Multimodal Machine Learning}
\newacronym{av}{AV}{Audio-Visual}
\newacronym{rl}{RL}{Representation Learning}
\newacronym{dl}{DL}{Deep Learning}
\newacronym{soa}{SOTA}{State-of-the-Art}
\newacronym{ed}{E-D}{Encoder-Decoder}

\newacronym{mfcc}{MFCC}{Mel Frequency Cepstral Coeficients}
\newacronym{dct}{DCT}{Discrete Cosine Transform}
\newacronym{lms}{LMS}{Log-Mel Spectrograms}

\newacronym{rnn}{RNN}{Recurrent Neural Networks}
\newacronym{lstm}{LSTM}{Long Short Term Memory Networks}

\newacronym{ar}{AR}{Action Recognition}
\newacronym{nn}{NN}{Neural Networks}

\newacronym{cca}{CCA}{Canonical Correlation Analysis}
\newacronym{ce}{CE}{Cross-Entropy}
\title{Recent Advances and Challenges in Deep Audio-Visual Correlation Learning}

\author{
Luís Vilaça$^{1}$\footnote{Luis was involved in this work during his internship from September 2021 to March 2022 in National Institute of Informatics (NII), Tokyo.}$^3$\and
Yi Yu$^{1}$\And
Paula Viana$^{23}$\\
\affiliations
$^1$National Institute of Informatics, Japan\\
$^2$Polytechnic of Porto, School of Engineering, Porto, Portugal\\
$^3$INESC TEC, Porto, Portugal\\
\emails
luis.m.salgado@inesctec.pt,
yiyu@nii.ac.jp,
paula.viana@inesctec.pt
} 

\begin{document}
	
	\maketitle
	
	\begin{abstract}
	Audio-visual correlation learning aims to capture essential correspondences and understand natural phenomena between audio and video. With the rapid growth of deep learning, an increasing amount of attention has been paid to this emerging research issue. Through the past few years, various methods and datasets have been proposed for audio-visual correlation learning, which motivate us to conclude a comprehensive survey. This survey paper focuses on \acrfull{soa} models used to learn correlations between audio and video, but also discusses some tasks of definition and paradigm applied in AI multimedia. In addition, we investigate some objective functions frequently used for optimizing audio-visual correlation learning models and discuss how audio-visual data is exploited in the optimization process. Most importantly, we provide an extensive comparison and summarization of the recent progress of \acrshort{soa} audio-visual correlation learning and discuss future research directions.\par
\end{abstract} 
	
	\section{Introduction}
	\label{section:introduction}
	
	A wide variety of multimedia information and data such as image, text, video, and audio is aggregated on the Internet over time, bringing opportunities to learn knowledge from their structure hidden in such kind of heterogeneous data through machine learning and artificial intelligence. Understanding knowledge and structure hidden in natural audio-visual phenomena requires being able to process multiple signals that compose the concept of what we are experiencing. The natural characteristics and temporal correlations in audio and video increase the complexity of capturing cross-modal information, requiring multimedia technology and machine learning methods able to extract these semantic sequential relations \cite{yu2012automatic}. However, their inconsistent distributions and representations cause a heterogeneous gap, which compromises the correlation of audio and video in common spaces \cite{shah2014advisor}.\par
	
	\begin{figure*}
		\centering
		\includegraphics[width=0.80\textwidth]{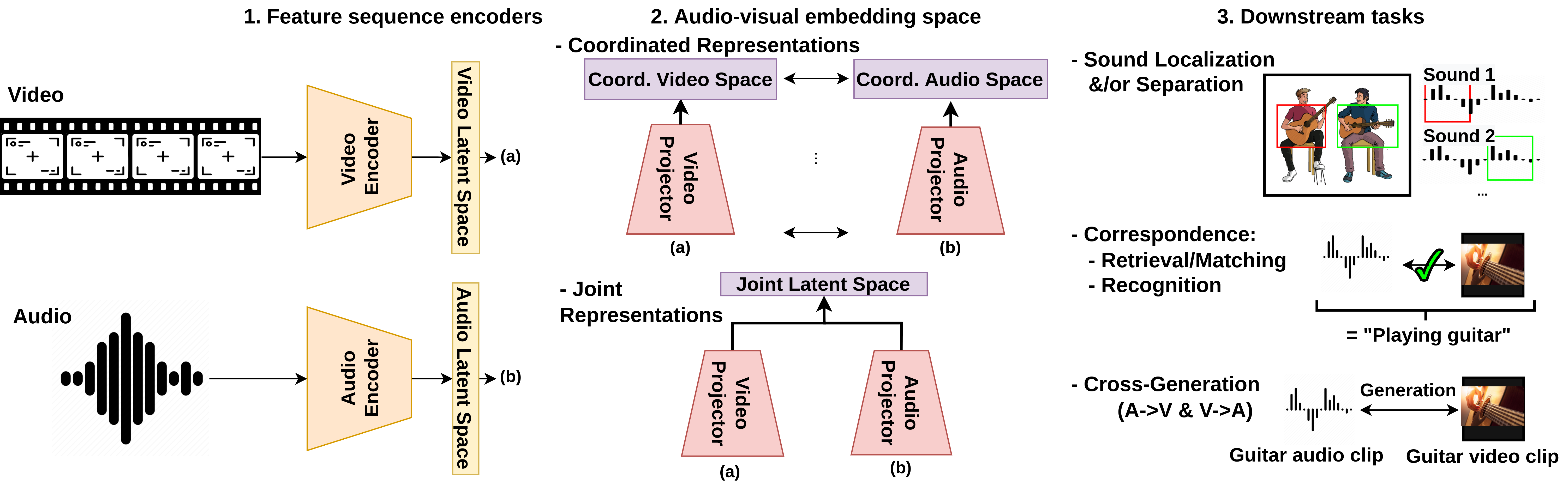}
		\caption{Audio-visual correlation learning and subsequent downstream tasks} 
		\label{fig:generic-approach}
	\end{figure*}
	
	The emergence of deep learning and available audio-visual datasets has facilitated the development of audio-visual correlation learning. Deep \acrfull{avcl} aims to embed audio and visual feature vectors through deep neural networks into a common space. All application scenarios propose to learn better representations for their specific goals by either constraining the outputs to have desired statistical properties or linking relevant features/sub-components between them to synchronize semantic contexts (Figure \ref{fig:generic-approach}) \cite{MViewSurvey,MMMLSurvey}. In fact, coordinating sub-spaces for different modalities through constrains, such as correlation (TNN-C-CCA \cite{zeng2020deep}), demonstrated excellent performances in downstream tasks such as cross-modal retrieval \cite{zeng2021learning}. In addition, in the \acrshort{soa}, adversarial learning, reconstruction or weighted pooling processes through models such as \acrfull{gan} \cite{zheng2021adversarial,seo2020hmtl}, \acrfull{ae} \cite{recasens2021broaden,zhu2021learning} and Attention mechanisms \cite{zheng2021deep,min2021cross} have been investigated in recent years to improve the capability of \acrshort{avcl} methods.\par
	
	Existing surveys introduce generic applications, core ideas and theoretical concepts of multimodal machine learning \cite{MViewSurvey,MMMLSurvey}, and introduce audio-visual applications in the multimedia field \cite{DAVLearningSurvey}. In contrast, this survey paper addresses the recent progress of deep learning models and their objective functions used for \acrshort{avcl}. We address the natural complementary relationship between vision and audio that makes them suitable for exploiting underlying semantics. By analyzing the latest works on the field, we identified that the proliferation of attention-based models and objective functions based on semi-supervised learning have provided a new bloom for the field as it can be observed by the high performances demonstrated in Table \ref{tab:methods-final}. Consequently, it is important to summarize new advances and discuss their pros and cons through a systematic comparison among existing deep \acrshort{avcl} approaches.\par
	
	\section{Problem Definition}
	\label{section:problem-definition}
	The problem of Deep \acrshort{avcl} can be formulated as follows. We are interested in data consisting of tuples ${x^{(n)} = (x_a^n, x_v^n)}_{n=1}^N$, where $x_a$ and $x_v$ are features extracted for audio and video respectively. Both representations belong to separate sub-spaces and can have different dimensions, thus we represent them as $x_a \in \mathbb{R}^{d_a}$ and $x_v \in \mathbb{R}^{d_v}$ for audio and video respectively. We seek to project both audio and video through \acrfull{dnn} into their own feature vector sub-spaces where the joint space composed by their aggregation preserves the specific cross-modal information contained in each modality. That is to say, these representations are projected into a common space $S$ through $f(x_a;W_{a})$ and $g(x_v;W_{v})$, which respectively provide the sub-spaces $S_{\text{audio}}$ and $S_{\text{video}}$. The weights of $f$ and $g$ are adjusted through back-propagation in order to maximize the correlation between audio and video.\par
	
	This problem can be established as a supervised, semi-supervised or unsupervised task depending on how relations between modalities are used. For instance, exploring relations within tuples $(x_{a}^{n_1}, x_{v}^{n_2})_{n_1 = n_2}$ and assuming each pair as having specific semantics allows to extract correlations in a semi-supervised way. In contrast, it can also be desirable to relate modalities as a whole $(x_{a}^{n_1}, x_{v}^{n_2})_{n_1 = n_2 \vee n_1 \neq n_2}$ and take into account supervised information (i.e., labels) regarding each tuple.\par
	
	\section{Learning Audio-Visual Correlation}
	\label{section:learningAVCL}
	
	In cross-modal correlation learning between audio and visual modalities , feature vectors of audio and visual sequences can be extracted directly by existing pre-trained models \cite{zhang2019deep,zeng2018audio} or feature extractions are trained by following several DNN layers \cite{zhen2019deep,zeng2021learning}. However, due to inconsistent distributions and representations between them, we have to learn a common space which can be utilized for bridging this gap, and further measuring and maximizing their correlation. In other words, since our input data ($x_a$ and $x_v$) belong to sub-spaces with different sizes, correlating audio-visual cross-modal data is done by using separate projections ($f(x_a; W_a)$ and $g(x_v; W_v)$). Generally, Deep \acrshort{avcl} contains mainly two steps: i) To extract the temporal information from the frame-level audio and visual feature sequences, $x_a$ and $x_v$ are fed into two separate sequence encoders. To better learn sequential dependencies between audio and visual data, recent works, such as attention mechanism, focus on synchronizing cross-modal semantic information and minimizing the heterogeneity gap simultaneously; ii) Objective functions are used to optimize these two sub-networks to maximize the correlation between audio and video in the joint space through back-propagation. According to different kinds of audio-visual data are exploited in deep learning stage, various objective functions are used in existing methods. In this section, we address the recent advances for step i) and ii) and discuss some tasks and paradigm in Deep \acrshort{avcl}. In Table \ref{tab:methods-final} we provide a brief summary of these \acrshort{soa} \acrshort{avcl} methods.\par
	
	\subsection{Sequence Encoding Models}
	\label{subsection:models}
	
	Various sequence encoding models are proposed to capture and synchronize semantic temporal information between audio and visual data, which facilitates to build the audio-visual embedding spaces.\par
	
	\subsubsection{Attention}
	\label{subsection:attention-model}

	In neural networks, the effect of attention aims to enhance some parts of the input data while enervating other parts. It can be utilized for enhancing semantic cross-modal information to align audio and visual sequences through a dynamic pooling process that allows to learn relations between sub-elements (i.e., segments of audio or visual locations). It consists in a dynamic weighted sum of vectors with scalar values (i.e., probabilities) obtained through an interaction/score function (e.g., addition, dot-product, scaled dot-product). Therefore, it allows to obtain context from different sources (within or between modalities) to encode a given input (e.g., using $x_a$ to encode $x_v$). We identified the following methodologies:\par

	\paragraph{\textit{i. Co-Attention}} enables the learning of pairwise attentions, which can be used to explore inter-modal correlations by encoding a feature vector with context from another. Typically, a residual connection is added in the attended feature vector, which can be seen as adding missing relevant information from one modality to another. Co-attention allows to learn correlations between multiple pairs of data \cite{zheng2021deep,duan2021audio}. Nevertheless, the context used to encode one representation can also be drawn from features extracted at multiple scales (i.e., different framerates) \cite{wu2019unified}, different timesteps \cite{ma2020contrastive} or between different sub-elements (e.g., image regions or patches) \cite{lin2021exploiting,gan2020music}. On a separate note, co-attention can also be used to coordinate/align the projectors' weights \cite{min2021cross}.\par

	\paragraph{\textit{ii. Self-Attention}} relates different positions of a sequence to calculate a representation of this sequence, which can be used to explore the sequential structure of each modality (inter-modal) and it can be combined with co-attention to weight intra and inter-modal relations \cite{cheng2020look}. It is the base of the \acrfull{tf} architecture which is applied in two different ways: multi-head self-attention (encoder) and masked self-attention (decoder). They both consist in the same process of adding several parallel self-attention projections. However, masked self-attention limits the context of each one to avoid attention bias (i.e., any query in the decoder only attends to all positions up to the current one).\par

	\paragraph{\textit{iii. \acrlong{tf}}} is an encoder-decoder model with self-attention mechanism as well as positional encoding, which can be exploited to capture sequential cross-modal semantic information between audio and visual cross-modal data. \acrshort{tf}s are formed by an encoder (i.e., multi-head), a decoder (i.e., masked) and an encoder-decoder module. The \acrshort{tf} encodes the output from the decoder using the context of the encoder. Similar to co-attention, the \acrshort{tf} encoder can be replaced by different feature extractors or features from sub-elements of each modality (e.g., past and future frames or different patches within the same frame) and leverage the implicit co-attention mechanism between modalities \cite{gan2020foley,lin2020audiovisual,morgado2020learning}. Furthermore, \acrshort{tf}s can also be used to relate cross-translations between modalities of the same media asset to align/correlate them in sequence-to-sequence generation \cite{morgado2020learning}.\par
	\paragraph{Discussion:} Attention mechanisms allow to fuse and coordinate data from different sub-spaces by selectively weighting the importance of each element in each one. In some scenarios, redundant/relevant information is respectively removed or added \cite{zhu2021leveraging,chen2021localizing,zhu2020visually}. The context used for weighting the representations can be obtained from different combinations or features of $x_a$ and $x_v$ or from their sub-elements. Thus, establishing where (and when) to draw context depends on the available data and the target application, but opens a pathway for many possible extensions based on attention for audio-visual alignment. Nevertheless, attention requires an interaction/score function that fits the constraint length of each sub-space and expresses the intended correlation between its inputs. For instance, addition requires inputs with equal length and accumulates the growth of both vectors, while the dot product can accept mixed lengths but capture the directional growth of one into the other. Therefore, defining the interaction function plays a crucial role for many applications when attention is exploited. On a final note, similarly to the attention mechanism (co and self-attention), \acrshort{tf}s can also be used to learn inter and intra-modal relations. Higher performances generally can be achieved as the parallel connections are associated with the masked and multi-head self-attention modules. However, it should also be noted that the number of parameters of these models is substantially large. In fact, some studies have already showed that the number of parameters should escalate with $N$ \cite{hahn2020theoretical} to achieve a perfect accuracy/cross-entropy on the elements in a sequence of length $N$.
	
	\subsubsection{Auto-Encoders}
	\label{subsubsection:aes}
	
	\acrshort{ae}s are unsupervised encoder-decoder models that create latent representations through reconstruction. The encoder $f_{\theta_1}(x)$ computes the latent space $h=f_{\theta_1}(x)$ from the input $x$ and the decoder computes its reverse mapping, $\hat{x}=g_{\theta_2}(h)$. Therefore, the latent representation reflects the structural distribution of the original data (similar to summarizing/reducing dimensionality). We identified the following methodologies:\par
	
	
	i. Using a shared latent representation: through concatenation of uni-modal latent spaces \cite{wang2018associative}; through attention between uni-modal latent spaces \cite{zhao2019sound}.\par
	ii. Injecting supervised semantic context using the latent space for classification \cite{rajan2021robust}\par
	iii. Imposing similarity constraints between separate latent spaces \cite{cao2016correlation}\par
	iv. Injecting additional information through attention in the latent spaces to condition the reconstruction \cite{gao2019co,lin2021exploiting}\par 
	v. Leveraging similarities between reconstructions and cross-reconstructions \cite{recasens2021broaden,wang2018associative,rajan2021robust}\par
	vi. \acrfull{vae} reconstruction conditioned by multiple inputs \cite{zhu2021learning,zhang2021variational}\par
	\paragraph{Discussion:} Through reconstruction, \acrshort{ae}s are an efficient way to reduce the dimensionality of input features while preserving the same distribution of the original data, which can be applied for creating individual latent spaces of the same size (can be used as projectors) and balancing the information between projected sub-spaces. For instance, cross-modal discrepancy can be reduced by \acrshort{ae}s, when the input features are obtained with different feature extractors (different compression levels). This helps to balance the results in applications where only one modality is given as input (e.g., cross-modal retrieval). Instead of strictly defined relations between modalities, using shared representations in reconstruction has the advantage of implicitly allowing each one to leverage insight from itself and from others. This results in having a lower heterogeneity gap when compared with an approach with separate latent spaces. We have seen consistently better results than traditional methods based on \acrfull{cca} and Deep \acrshort{cca} \cite{wang2018associative}. Moreover, shared latent spaces make the input share the same context and apply constraints that affect all modalities.\par
	
	\subsubsection{Generative Adversarial Networks}
	\label{subsubsection:gan-model}
	
	Generative Adversarial Networks (GAN) are composed by two components: a generator $G$ and a discriminator $D$. The generator $G$ learns to produce target samples as similar to the real ones as possible to confuse the discriminator $D$, that attempts to distinguish generated samples from the real ones, keeping itself from being confused. Instead of directly mapping a latent representation from the data as \acrshort{ae}s, \acrshort{gan}s learn to implicitly map latent random representations to ``samples belonging to the training set'', which try to narrow the difference between distributions of real (i.e., training set) and generated data. We identified the following methodologies:\par
	
	i. Heterogeneity gap minimization through adversarial learning \cite{zheng2021adversarial,seo2020hmtl}\par
	ii. Using GANs and AEs for cross-modal generation \cite{mira2021end,fanzeres2021sound,athanasiadis2020audio}\par
	iii. Condition the discriminator by injecting context semantics in the form of: Class-labels \cite{athanasiadis2020audio}; Latent representations of other modalities \cite{fanzeres2021sound}; Classifiers' prediction on generated samples \cite{athanasiadis2020audio};\par
	iv. Joint optimization between adversarial and other loss functions \cite{seo2020hmtl,mira2021end}\par
	
	\paragraph{Discussion:} For audio-visual cross-modal retrieval, the standard \acrshort{gan}s adversarial scheme can be used to minimize the heterogeneity gap and obtain modality-independent representations \cite{zheng2021adversarial} or to transfer knowledge between modalities \cite{seo2020hmtl}. These methods are often complemented with deep metric learning frameworks \cite{zheng2021adversarial} or classification \cite{seo2020hmtl} to include semantic context. In contrast, from a generation point-of-view, we can also leverage reconstruction. The random vector is replaced by latent representations obtained from a model, typically an encoder-decoder, used for generating one modality with the other \cite{mira2021end,fanzeres2021sound,athanasiadis2020audio}. However, in this case the main differences lie in how complementary information is explored. The discriminator receives an input and tries to classify it into real or not, but its decision can be conditioned by adding information in the last layer before the classification. In addition, these models can also consider several sources of conditional information: latent representation (from \acrshort{ae}s) \cite{fanzeres2021sound} or class-label information.\par
	

	
\begin{table*}[!ht]
\centering
\caption{Overview of representative works for audio-visual correlation learning}
\label{tab:methods-final}
\resizebox{\textwidth}{!}{%
\begin{tabular}{lllllll}
\hline
Paper & Objective Function & Learning & Application & Metrics & Performance & Benchmark Datasets \\ \hline
\cite{fanzeres2021sound} & MSE + Adv.L. & Semi-Supervised & Sound-2-Image Generation & I & 0.14 & VEGAS \\
\cite{zhu2021learning} & ELBO + Wasserstein & Unsupervised & Cross-Modal (AV) Generation & MRR & 0.175 & AVE \\
\cite{mira2021end} & Adv.L & Unsupervised & Video-2-Speech Generation & WER & 23.1/42.5 & GRID/LRW \\
\cite{gan2020foley} & CE & Supervised & Music Generation & NDB & 20 & MUSIC \\
\cite{athanasiadis2020audio} & Adv.L & Unsupervised & Audio \& Video Generation & FID & 49.8/59.4 & CREMA-D/RAVDESS \\
\cite{rajan2021robust} & MSE + DCCA + CE & Supervised & Emotion Recognition & ACC & 56.5(A)/55.2(V) & SEW \\
\cite{seo2020hmtl} & Adv.L + CE & Semi-Supervised & Emotion Recognition & ACC & 65.2/40.1 & CMU-MOSI/IEMOCAP \\
\cite{zhang2019deep} & CE & Supervised & Emotion Recognition & ACC & 62.5 & AFEW8.0 \\
\cite{recasens2021broaden} & L2 & Semi-Supervised & Action Recognition/Sound Classification & ACC & 93.2/69.9/93. & UCF101/HMDB51/ESC-50 \\
\cite{ma2020active} & NCE & Semi-Supervised & Action Recognition/Sound Classification & ACC & 94.1/67.2/90.8 & UCF101/HMDB51/ESC-50 \\
\cite{min2021cross} & NCE & Semi-Supervised & Action Recognition/Sound Classification & ACC & 90.3/61.1/81.4 & UCF101/HMDB51/ESC-50 \\
\cite{akbari2021vatt} & NCE & Semi-Supervised & Action Recognition/Sound Classification & ACC & 89.6/65.2/84.7 & UCF101/HMDB51/ESC-50 \\
\cite{cheng2020look} & CE & Semi-Supervised & Action Recognition & ACC & 87.8/58.2 & UCF101/HMDB51 \\
\cite{rouditchenko2020avlnet} & MMS & Semi-Supervised & Text-2-Video Retrieval & Recall@10 & 49.9/67.9 & MSR-VTT/YouCook2 \\
\cite{patrick2020multi} & NCE & Semi-Supervised & Sound Classification/Action Recognition & ACC & 88.5/95.2/72.8 & ESC-50/UCF101/HMDB51 \\
\cite{MMVAudioSet} & NCE + MIL-NCE & Semi-Supervised & Sound Classification/Action Recognition & ACC & 88.9/91.8/67.1 & ESC-50/UCF101/HMDB51 \\
\cite{afouras2021self} & NCE/CE & Supervised & Cross-Modal (AV) Retrieval & mAP@50 & 39.4/28. & VGG Sound/AudioSet \\
\cite{pretet2021cross} & T & Semi-Supervised & Cross-Modal (AV) Retrieval & Recall@10 & 7.425 & MVD \\
\cite{zheng2021adversarial} & LSL + Adv.L & Supervised & Cross-Modal (AV) Retrieval & ACC & 93. & VGGFace + VoxCeleb \\
\cite{zeng2020deep} & T & Supervised & Cross-Modal (AV) Retrieval & mAP & 74.2 & VEGAS \\
\cite{merkx2019language} & Hinge & Supervised & Speech-2-Image Retrieval & Recall@10 & 52.3 & Flikr-8K \\
\cite{ilharco2019large} & MMS/T & Supervised & Cross-Modal (AV) Retrieval & Recall@10 & 52.65 & Flickr Audio Caption Corpus \\
\cite{he2019new} & CE/L1/QL & Supervised & Cross-Modal (AVT) Retrieval & mAP & 41.2 & CUB-200-2011 + Youtube Birds \\
\cite{horiguchi2018face} & NPL & Unsupervised & Face-2-Voice Retrieval & mAP & 2.07 & FVceleb \\
\cite{wu2019unified} & L2 & Supervised & Text-2-Image Retrieval & Recall@10 & 71.9 & MS-COCO \\
\cite{cao2016correlation} & L2 & Supervised & Cross-Modal (IT) Retrieval & mAP & 35.5/71.7 & Wiki Dataset/Flickr \\
\cite{zhang2021variational} & ELBO + Corr. + Center + L2 & Supervised & Cross-Modal (AV) Retrieval & mAP & 81.2/35. & VEGAS/AVE \\
\cite{zhen2019deep} & CE + Center + L2 & Supervised & Cross-Modal (IT) Retrieval & mAP & 71.6/61.3 & Pascal Sentence/NUS-WIDE-10k \\
\cite{zeng2021learning} & Corr. + L2 + Frob. Norm & Supervised & Cross-Modal (AV) Retrieval & mAP & 77.8/30.8 & VEGAS/AVE \\
\cite{chen2021localizing} & NCE & Semi-Supervised & Sound Localization & AUC & 0.573/0.590 & Flickr-SoundNet/VGG-SS \\
\cite{lin2020audiovisual} & CE & Semi-Supervised & Sound Localization & ACC & 76.8 & AVE \\
\cite{morgado2020learning} & NCE & Semi-Supervised & Sound Localization & ACC & 73.85/73.8/37.66 & YT-360/UCF-101/HMDB51 \\
\cite{ramaswamy2020see} & CE/T & Semi-/(Un)Supervised & Sound Localization & ACC & 74.8 & AVE \\
\cite{duan2021audio} & CE & Supervised & Sound Localization & ACC & 76.2 & AVE \\
\cite{lin2021exploiting} & CE & Semi-Supervised & Sound Separation & STFT & 0.865/1.448 & FAIR PLAY/YT-MUSIC \\
\cite{zhu2020visually} & CE & Supervised & Sound Separation & SDR & 8.25 & MUSIC \\
\cite{gao2019co} & L1 + CE & Supervised & Sound Separation & SDR & 7.64 & MUSIC \\
\cite{zhao2019sound} & CE & Semi-Supervised & Sound Separation & SDR & 8.31 & MUSIC \\
\cite{zhu2021leveraging} & CE & Semi-Supervised & Sound Separation \& Localization & SDR & 10.74 & MUSIC \\
\cite{hu2020curriculum} & CL & Semi-Supervised & Sound Separation \& Localization & AUC/SDR & 49.2/6.59 & Flickr-SoundNet/MUSIC \\ \hline
\end{tabular}%
}
\end{table*}

	\subsection{Objective Functions}
	\label{section:learning-frameworks}
	
	Objective functions are used to adjust audio-visual embedding models through back-propagation to maximize the correlation between modalities in the common space. This can help to maintain and capture specific audio-visual semantics. Typically,  several regularization terms are associated to constrain the embeddings in order to mitigate the heterogeneity gap and to increase correlation.\par
	
	\paragraph{\textit{i. \acrfull{ce}}} and other distance-based extensions can be used \cite{he2019new,zhen2019deep} to achieve inter-class discrimination via a projection to a latent semantic sub-space. \acrshort{ce} uses entropy as a measure of distance between the true and the predicted distribution. In contrast, distance-based works use distance measures between joint spaces with the same size as the one-hot encoded vectors and minimize the distance between the \acrshort{nn} output and the ground truth \cite{zhen2019deep}.\par
	
	\paragraph{Discussion:} Despite being able to establish relations between semantics, they do not enforce any intra-class relation. To ensure intra-class consistency, some proposals use the center loss as regularization (L1 distance between the center of each class and each uni-modal embedding) \cite{zhen2019deep,he2019new}. Similarly, to align both modalities and reduce the heterogeneity gap other works constrain them to have similar representations using the distance between embeddings (e.g., L2, cosine similarity) \cite{min2021cross}. All these terms are typically leveraged in a joint optimization scheme for \acrshort{avcl}.\par
	
	\paragraph{\textit{ii. Hinge Loss (and extensions)}} are used in several works to increase compactness and separability at the same time, because traditional \textit{softmax} losses cannot directly enforce them. Given a similarity function between embeddings (e.g., cosine similarity), it imposes a margin between correctly classified and misclassified samples. Comparing positive and negative relationships is allowed between embeddings, where the negative samples are drawn from each mini-batch and the margin controls its amount. This approach is able to improve the capability of correlation learning by exploiting the comparison between different modalities of classified and misclassified samples. Margins can be fixed (hinge loss) to establish a fixed distance between the similarity of matched pairs and mismatched ones. It is computed between two pairs of embeddings as follows:
	\begin{align}
	L(c, i, \theta) = \sum_{(c,i) \in \beta} ( \max(0, \cos(c,i) + \alpha) + \nonumber\\ \max(0, \cos(c,i) + \alpha) ),
	\end{align}
	where $i,c$ are pairs with different modalities (i.e., audio and video from the same clip) \cite{merkx2019language}. $cos()$ represents the cosine similarity between embeddings but it can be replaced by other similarity function.\par
	
	Because the relationships between embeddings change during training, margins can be progressively adapted to fit those changes. Therefore, \cite{ilharco2019large,rouditchenko2020avlnet} proposed to use monotonically increasing margins (i.e., Masked Margin Softmax (MMS)). Works that leverage MMS typically use it bidirectionally between audio and video. Therefore, using a mini-batch of size $B$, it is computed as follows:\par
	\begin{align}
	L_{mms} = L_{av} + L_{va}, \nonumber\\
	L_{av} = -\frac{1}{B} \sum_{i=1}^B \log [\frac{e^{\cos(ii) - \sigma}}{e^{\cos(ii) - \sigma} + \sum_{i=1}^B M_{ij}e^{\cos(ij)}}], \nonumber\\
	L_{va} = -\frac{1}{B} \sum_{i=1}^B \log [\frac{e^{\cos(jj) - \sigma}}{e^{\cos(jj) - \sigma} + \sum_{i=1}^B M_{ij}e^{\cos(ij)}}],
	\end{align}
	where $M_{ij}$ controls if both embeddings are positively associated or not (masking). The similarity between embeddings ($\cos$) is subject to a margin $\sigma$, scheduled to change during training according to a predefined scheduler. Nevertheless, margins can also adapt to the current mini-batch by using the mean distance between positive and negative pairs \cite{monfort2021spoken}.\par
	
	\paragraph{Discussion:} In sum, margin-based approaches for \acrshort{avcl} allow to sample negative pairs efficiently (randomly drawn), but the initial margin values must be carefully tweaked to avoid divergence. Moreover, dynamic margin heuristics can alleviate the issue of under or oversampling negative pairs but at the cost of creating different convergence rates for similar models.\par
	
	\paragraph{\textit{iii. Triplet loss (TL) (and extensions)}} is calculated with three samples at each training step (anchor - input, positive and negative). Using a projector with shared weights (for positive and negative samples), it aims to maximize the distance between the anchor (input) and the negative while minimizing its distance to the positive sample. It is computed as follows:\par
	\begin{align}
	L_{triplet}(x,x^+,x^-) = \sum_{x \in \mathcal{X}} \max(0,d_{pos}-d_{neg}+\zeta),\nonumber\\
	d_{neg} = ||f(x)-f(x^-)||^2_2, d_{pos} = ||f(x)-f(x^+)||^2_2,
	\end{align}
	where $\zeta$ is the minimum distance between the relations of matched and mismatched pairs. For a fair convergence, this approach requires to select challenging negative samples, which slows training and increases computational demands.\par
	
	The combination of positive and negative samples can be done in several ways. \cite{ramaswamy2020see} gathered only positives from the same media asset. Similarly, \cite{pretet2021cross} used it bidirectionally by leveraging each modality as an anchor separately and jointly optimizing the model. By using distances between different modalities, the authors were able to learn multimodal correlations and align both spaces. In contrast, the triplet loss can also be used to leverage supervised information (using class labels as reference) from different modalities in order to maximize semantic discrimination \cite{zeng2020deep}.\par
	
	\paragraph{Discussion:} The triplet loss is limited to one to one relationships between samples. For this reason, several extensions were proposed to generalize the comparison with multiple negative or positive samples \cite{horiguchi2018face,he2019new} (Lifted Structured Loss (LSL), N-Pair Loss (NPL) and Quadruple Loss (QL)). However, they still have the burden of requiring a methodology for selecting the best negative candidates.\par
	
	\paragraph{\textit{iv. Contrastive Loss (and extensions)}} takes a pair of inputs and minimizes the distance of samples from the same class, maximizing it otherwise. In contrast to triplet loss and margin-based approaches, the negative samples are usually obtained at random which reduces its dependency on negative sampling.\par
	
	\begin{align}
	L_{contr}(x_a, x_v, \Theta) = 1[y_i = y_j] \cos(x_a, x_v)^2 + \nonumber\\+ 1[y_i \neq y_j]\max(0, \zeta - \cos(x_a, x_v))^2,
	\end{align}
	where $\zeta$ establishes a margin between samples of different classes. It can be viewed as an extension of margin-based loss functions but the margin is only applied between positive and negative samples. However, in contrast to triplet loss, it only considers one pairwise relation at each training step.\par
	
	Similarly to previous approaches, for audio and video we identified that negative samples can be obtained in an semi-supervised way by using paired modalities to learn correlations/alignment (i.e., only pairing audio and video when they describe the same media asset) \cite{hu2020curriculum}. In addition, supervised information can also be leveraged bidirectionally from audio and video embeddings with the same semantics for ensuring semantic discrimination (i.e., from the same class) \cite{ma2020active}.\par
	
	\paragraph{Discussion:} According to the application scenario, negative and positive candidates can be obtained in a variety of ways. For instance, for learning correlations between spatial locations, \cite{chen2021localizing} proposed to contrast image regions with audio segments to learn the association between a given sound and its visual representation. Similarly, for learning temporal alignment between events \cite{patrick2020multi} used the augmented versions of both modalities (e.g., temporal shifts, augmentations in the embeddings). In sum, depending on the application and how pairs of samples are generated, the contrastive objective function can be used for learning the alignment/correlation between different modalities. Alternatively, semantic discrimination is added to the representations when this information is available.\par
	
	\paragraph{\textit{v. Noise Contrastive Estimation (NCE) and extensions}} are a set of contrastive learning objectives used to distinguish positive and negative pairs of embeddings. Instead of using the original NCE function, current \acrshort{soa} methods use a surrogate version that maximizes mutual information (InfoNCE). Assuming the input embedding $x_a$ (audio in this example) and a set of $N$ negative samples (usually taken from the opposing modality for alignment purposes - video), infoNCE is computed as:
	\begin{align}
	L_{nce}(x_a, x_v) = - \log \frac{e^{\frac{\cos(x_a,x_v)}{\tau}} }{e^{\frac{\cos(x_a,x_v)}{\tau}} + \sum_{x'\in N} e^{\frac{\cos(x_a,x_v)}{\tau}}},
	\end{align}
	where N is the set of negative modality pairs for audio and $\tau$ the temperature parameter to control the concentration of features in the representation space (it controls the contribution between small and large distances). For example, this framework allows the comparison with multiple negative elements and extensions of NCE typically include the comparison with multiple positive instance (Multiple Instance MIL-NCE) \cite{akbari2021vatt,MMVAudioSet}.\par
	
	\paragraph{Discussion:} Temporal co-occurrence is explored between modalities to increase their correlation. \cite{akbari2021vatt} and \cite{afouras2021self} leveraged positive samples from paired modality data and contrasted them with others. This has proven superior performance when compared with the approaches for learning alignments/correlations. In fact, representations extracted using NCE (and its extensions) reported top-1 accuracy in several benchmark datasets for audio-visual classification \cite{rouditchenko2020avlnet,akbari2021vatt}. However, NCE is mainly used for alignment between modalities and semantic discrimination is left to a posterior phase (as a downstream task) \cite{afouras2021self}.\par

	\section{Conclusion}
	\label{section:conclusion}
	
	As reflected in Table \ref{tab:methods-final}, recent researches have intensively demonstrated that i) semi-supervised learning methods are able to achieve more discriminative features as it allows to leverage large volumes of (unlabeled) data, and ii) attention-based methods are designed to improve the synchronization between audio and visual sequences.\par
	
	 The typical scenarios for semi-supervised learning frameworks loose the definition of labels and create negative and positive samples in supervised or semi-supervised ways. Semi-supervised learning requires very large batch size, specially for methods that use in-batch samples (e.g., margin-based losses). This allows them to obtain diverse representations of the data at each step, while covering a wide range of negative samples and enabling the model to learn meaningful representations. However, the time complexity of model training will increase with the batch size. Consequently, their computational requirements are very large, which hinders the democratization of these AVCL methods. Furthermore, the performances of semi-supervised learning methods are limited in evaluating classification-based downstream tasks. It is necessary to apply semi-supervised learning approaches to a wider range of cross-modal tasks. Simultaneously, various attention mechanism can be developed to learn and understand the intra-modality and inter-modality correlation between audio and visual data, which helps to enhance fine-grained alignment and increase the performance of models. Moreover, attention along with memory networks can tackle long-term temporal signals for better sequence prediction.\par

	\section{Research Challenges and Directions}
	\label{section:challenges}
		
	As we know, cross-modal correlation learning aims to capture inter-modality information which is able to complement each other between different modalities. However, it is challenging to learn an ideal unified representation of paired audio-visual sequences that can not only extract semantic inter-modality information but also preserve intra-modality structures. Considering local or global sequence embedding methods with deep learning is promising to improve the capability of cross-modal correlation learning. Recent research progresses also have shown that unsupervised learning methods are able to achieve more discriminative features when trained on enough data, which have encouraged us to pretrain a deep model like VisualBERT on a very large amount of data in an unsupervised way and then fine-tune this model on much smaller datasets to realize specific tasks. Pre-trained models are also very important for multimodal downstream tasks. However, it remains unexplored in dealing with audio and visual sequences. Investigation of standard settings for multimodal audio-visual pre-trained models is indispensable for the research community, which will be very helpful for obtaining more knowledge about pre-trained mechanism.\par
	
	In addition, interpretability and reliability are very important properties of a human-level AI system. However, most of audio-visual correlation learning models still have serious problems such as how to explain the predicted result and showing how much the result is trustable. Therefore, it is crucial to develop interpretable cross-modal correlation learning methods to better extract and understand the advanced characteristics and correlations of audio and visual sequence data, and enable to explain when a model works and why a model might fail, and eventually increase the reliability of multimodal intelligence in various applied environments.\par

	\section*{Acknowledgements} This work was partially supported by NORTE 2020, under the PORTUGAL 2020 Partnership Agreement, through the ERDF within project NORTE-01-0145-FEDER-000065 (DECARBONIZE), and partially supported by JSPS Scientific Research (C) under Grant No.19K11987.

	\bibliographystyle{named}
	\bibliography{ijcai22}
	
\end{document}